\documentclass[aps,pra,twocolumn,showpacs,floatfix,nofootinbib,groupedaddress,superscriptaddress]{revtex4}
\usepackage{amssymb,graphics,graphicx,times,bm,bbm,multirow,color}

\begin{document}

\title{Environment-Assisted Quantum Walks in Photosynthetic Energy Transfer}
\author{Masoud Mohseni}
\affiliation{Department of Chemistry and Chemical Biology, Harvard University, 12 Oxford
St., Cambridge, MA 02138}
\author{Patrick Rebentrost}
\affiliation{Department of Chemistry and Chemical Biology, Harvard University, 12 Oxford
St., Cambridge, MA 02138}
\author{Seth Lloyd}
\affiliation{Department of Mechanical Engineering, Massachusetts Institute of Technology,
77 Massachusetts Avenue, Cambridge MA 02139}
\author{Al\'an Aspuru-Guzik}
\affiliation{Department of Chemistry and Chemical Biology, Harvard
University, 12 Oxford St., Cambridge, MA 02138}
\email{aspuru@chemistry.harvard.edu} 


\begin{abstract}
Energy transfer within photosynthetic systems can display quantum
effects such as delocalized excitonic transport. Recently, direct
evidence of long-lived coherence has been experimentally
demonstrated for the dynamics of the Fenna-Matthews-Olson (FMO)
protein complex [Engel \textit{et al.}, Nature 446, 782 (2007)].
However, the relevance of quantum dynamical processes to the exciton
transfer efficiency is to a large extent unknown. Here, we develop a
theoretical framework for studying the role of quantum interference
effects in energy transfer dynamics of molecular arrays interacting
with a thermal bath within the Lindblad formalism. To this end, we
generalize continuous-time quantum walks to non-unitary and
temperature-dependent dynamics in Liouville space derived from a
microscopic Hamiltonian. Different physical effects of coherence and
decoherence processes are explored via a universal measure for the
energy transfer efficiency and its susceptibility. In particular, we
demonstrate that for the FMO complex an effective interplay between
free Hamiltonian and thermal fluctuations in the environment leads
to a substantial increase in energy transfer efficiency from about
70\% to 99\%.
\end{abstract}

\volumeyear{year}
\volumenumber{number}
\issuenumber{number}
\eid{identifier}
\date{May 23, 2008}
\startpage{1}
\maketitle

\section{Introduction}

Photosynthesis is the natural mechanism for the capture and storage
of energy from sunlight by living organisms. Excitation energy is
absorbed by pigments in the photosynthetic antennae and subsequently
transferred to a reaction center where an electron-transfer event
initiates the process of biochemical energy conversion. In certain
bacterial systems and higher
plants light harvesting efficiency is indeed above 99\% \cite{Blankenship02}%
. Although this phenomenon has been studied for decades \cite%
{Damjanovi97,Ritz01,Grondelle04,Cho05}, a full description of the
underlying mechanism leading to this remarkably high efficiency is
yet not available. It has already been demonstrated experimentally
that the excitation energy transfer within chromophoric arrays of
photosynthetic complexes could involve quantum coherence under
certain physical conditions. In particular, this phenomenon has been
observed via electronic spectroscopy of delocalized exciton states
of light-harvesting complexes \cite{Damjanovi97,Grondelle04} and the
Fenna-Matthews-Olson (FMO) protein complex \cite{Cho05}.

The energy transfer mechanism in multichromophoric arrays can often
be described by a semiclassical F\"{o}rster theory which involves
incoherent hopping of the excitations between energy levels \cite%
{Forster65,Scholes03,MayBook}. In this method, the Coulomb
interaction among different sites is treated perturbatively to
calculate the probability of exciton hopping. The more general
approach for including coherent effects is given by Redfield theory
which provides a microscopic description of excitation dynamics via
a master equation in a reduced space of excitons in the weak phonon
coupling and Born-Markov approximation \cite{Redfield65}. An
equivalent approach to Redfield theory for calculating the diffusion
constant of excitons was proposed by Silbey and Grover \cite{GroverSilbey71}%
. Alternative methods to explore coherent and incoherent exciton
transfer were also introduced using a stochastic model (Haken and
Strobl \cite{HakenStrobl73}), and a generalized master equation
formalism (Kenkre and Knox \cite{KenkreKnox74,KenkreBook82}).

In order to study the nonlinear spectroscopy of molecular
aggregates, Zhang \textit{et al.} \cite{ZhangMukamel98} introduced a
modified Redfield equation for statically disordered exciton systems
which treats the diagonal elements of exciton-bath coupling in a
non-perturbative fashion. This approach was later used to model
energy transfer dynamics in light-harvesting complexes of higher
plants \cite{Grondelle04}. Yang and Fleming provide a comprehensive
comparison of F\"{o}rster, standard Redfield, and modified Redfield
theories in Ref. \cite{YangFleming02}. A generalized theory for
multichromophoric F\"{o}rster resonance energy transfer which
includes coherence effects within donors and acceptors, while
considering donor-acceptor interactions according to the standard F\"{o}%
rster model was also proposed by Jang, Newton, and Silbey \cite%
{Jang04,Cheng06,Jang07}. In another study, the effects of geometry
and trapping on energy transfer were examined in simple chromophoric
arrays within the Haken-Strobl model \cite{GaabBardeen04}. Recently,
direct evidence of quantum coherence in the dynamics of energy
transfer has been observed experimentally in the FMO complex
\cite{Engel07} and also in the reaction center of purple bacteria
\cite{Lee07}. These previous studies led us to further explore and
characterize quantum interference, decoherence effects, and their
interplay within the dynamics of photosynthetic complexes as
potential mechanisms for the enhancement of the energy transfer
efficiency. Here, we develop a quantum walk approach, based on a
quantum trajectory picture in the Born-Markov and secular
approximations, as a natural framework for incorporating quantum
dynamical effects in energy transfer, as opposed to a classical
random walk picture that can effectively describe the excitation
hopping in the F\"{o}rster model.

The concept of quantum walks originated by Feynman works in
connection with diffusion in quantum dynamics, in particular to
model the dynamics of a quantum particle on a lattice
\cite{Feynman64}, and also path integral formalism for discretizing
the Dirac equation \cite{Feynman65}. Continuous-time quantum walks
were also used by Klafter and Silbey to find
hopping time distribution functions in exciton dynamics \cite%
{KlafterSilbey80}. The formal discrete and continuous-time
approaches to
quantum walks\ were developed later, e.g., see Refs. \cite%
{Aharonov93,Kempe03}, including some investigations of model
decoherence effects \cite{Kendon06}. Purely \textit{unitary}
continuous-time approaches to quantum walks were employed in the
context of quantum information science, where they yield potential
exponential speedups over classical algorithms
\cite{Farhi98,Childs02}. The quantum walks are of particular
interest as potential computational tools \cite{Aharonov01}, and
applications to quantum cellular automata \cite{Meyer96}, quantum
optical systems \cite{Sandars02}, and coherent excitation transport
\cite{Blumen06}.

In this work, we develop a theoretical framework for studying the
role of quantum coherence in energy transfer dynamics in molecular
systems within the Born-Markov approximation in the Lindblad
formalism. Our approach is essentially equivalent to a Redfield
theory with the secular approximation. However, our
approach naturally leads to quantum trajectory picture in a
fixed-excitation reduced Hilbert space that can be described by the
concept of directed quantum walks in Liouville space. Quantum walks
in actual physical systems differ from idealized models of quantum
walks in several significant ways. First, Hamiltonians of physical
systems typically possess energy mismatches between sites that lead
to Anderson localization \cite{Anderson58}. Second, actual quantum
walks are subject to relatively high levels of environment-induced
noise and decoherence. The key result of this paper is that the
interplay between the coherent dynamics of the system and the
incoherent action of the environment can lead to significantly
greater transport efficiency than coherent dynamics on its own. We
introduce the concepts of energy transfer efficiency (ETE) and its
susceptibility and robustness and explore the dynamical effects of
coherent evolution and environmental effects at various temperatures
from a microscopic Hamiltonian formalism. For the FMO protein, we
show that a Grover-type quantum search \cite{Grover97} cannot
explain the high ETE of this complex. However, we demonstrate that a
directed quantum walk approach can be used for studying the energy
transfer efficiency as a function of temperature, reorganization
energy, trapping rate, and quantum jumps from sites to sites.
Moreover, we explore similar dependencies for the susceptibilities
of ETE with respect to basic processes contributing to the FMO
dynamics including the free Hamiltonian, the phonon bath\ jumps,
dephasing in the energy basis, transfer to the acceptor, and exciton
decay. We demonstrate that the efficiency increases from 70\% for a
purely unitary quantum walk to 99\% in the presence of
environment-assisted quantum jumps.

This article is organized as follows. In Sec. \ref{Lindblad},\ we
develop a Lindblad master equation in the site basis for studying
energy transfer of multichromophoric channel systems in the
Born-Markov approximation. In Sec. \ref{QW}, we introduce a quantum
walks formalism in Liouville space to describe the energy transfer
pathways. The definition of ETE is presented in Sec. \ref{ETE}. In
Sec. \ref{FMO}, we apply our theoretical approach for studying the
dynamics of FMO complex. Some concluding remarks are given in Sec.
\ref{Conclusion}.

\section{Lindblad master equation for multichromophoric systems}

\label{Lindblad}

The Fenna-Matthews-Olson protein acts as an energy transfer channel
in the biological process of photosynthesis connecting the base
plate of the antenna complex to the reaction center of green sulfur
bacteria. This type of functional role of an interacting
multichromophoric system can be formalized by the Hamiltonian for an
consisting of $N_{D}$ donors, $N_{C}$ channel chromophores, and
$N_{A}$ acceptors as:
\begin{equation}
H_{S}=\sum_{m=1}^{N}\epsilon _{m}a_{m}^{\dagger
}a_{m}+\sum_{n<m}^{N}V_{mn}(a_{m}^{\dagger }a_{n}+a_{n}^{\dagger
}a_{m}). \label{GeneralFreeHamiltonian-Eq}
\end{equation}%
The $a_{m}^{\dagger }$ and $a_{m}$ are the creation and annihilation
operators for an electron-hole pair (exciton) at chromophore $m$ and $%
\epsilon _{m}$ are the site energies (not including the BChl a
transition frequency $12500$\textrm{cm}$^{-1}$) and
$N=N_{D}+N_{C}+N_{A}$. The $V_{mn}$ are Coulomb couplings of the
transition densities of the chromophores, often taken to be of the
F\"{o}rster dipole-dipole form, $V_{mn}\sim \frac{1}{R_{mn}^{3}}(\mathbf{\mu }_{m}\cdot \mathbf{\mu }%
_{n}-\frac{3}{R_{mn}^{2}}(\mathbf{\mu }_{m}\cdot \mathbf{R}_{mn})(\mathbf{%
\mu }_{m}\cdot \mathbf{R}_{mn})),$ with $%
\mathbf{R}_{mn}$ the distance between site $m$ and $n$ and
$\mathbf{\mu }_{m} $ the transition dipole moment of chromophore $m$
\cite{Damjanovi97}. Note that in systems where chromophores are
closely packed (e.g., the FMO complex of green sulfur bacteria
\cite{Engel07}) or the site energies are (almost)
resonant (e.g., the LH1 ring of purple bacteria \cite{Damjanovi97}), $%
\epsilon _{m}$ can be of the same order of magnitude as $V_{mn}.$
Such cases
require a non-perturbative treatment of the coupling. In this work, the $%
V_{mn}$ include all intra-donor/channel/acceptor couplings and
inter-chromophoric couplings for donor-channel and channel-acceptor.
Here, we ignore the $V_{mn}$ for inter donor-acceptor coupling due
to large spatial separation, as for example chlorosomes and reaction
center in the green sulfur bacteria \cite{Li97}. Thus, excitation
transfer from donor to acceptor always occurs via the channel. We
also assume that for
donor-channel and channel-acceptor coherent couplings are weak, i.e., $%
V_{mn}\ll \epsilon _{m}$ for couplings into and out of the channel.
This implies that the energy transfer to and from the channel can be
described by semi-classical F\"{o}rster theory.

In this work, we study the role of the FMO protein acting as an
energy transfer channel in green sulfur bacteria. Thus, we focus on
the dynamics of a chromophoric channel of $N_{C}$ sites with
denoting the free Hamiltonian in the reduced channel Hilbert space
as $H_{C}.$ The Hamiltonian $H_{C}$ is
formally equivalent to the Hamiltonian of Eq.~(\ref%
{GeneralFreeHamiltonian-Eq}) with $N=N_{C}.$ We consider only the
zero and single excitation manifolds given by the states $|0\rangle
$ and $|m\rangle
=a_{m}^{\dagger }|0\rangle $. We denote the eigenbasis of the Hamiltonian $%
H_{C}$ as exciton basis $|M\rangle =\sum_{m}c_{m}(M)|m\rangle $, where $%
H_{C}|M\rangle =\epsilon _{M}|M\rangle $. Here, the effect of the
donor is modeled by a static initialization of the channel. To
account for the channel-acceptor coupling we introduce an effective
non-Hermitian channel-acceptor Hamiltonian, $-iH_{C\rightarrow
A}=-i\sum_{m=1}^{N_{C}}\kappa _{m}a_{m}^{\dagger }a_{m}$, which can
be obtained by a projector-operator method analogous to Ref. \cite{MukamelBook}%
, with the acceptor transfer rates $\kappa _{m}=2\pi \int d\epsilon
_{a}\left\vert V_{ma}\right\vert ^{2}\delta (\epsilon _{m}-\epsilon
_{a})D(\epsilon _{a}),$ where $D(\epsilon _{a})$ denotes the density
of states for the acceptor.

In general, the multichromophoric channel is subject to a thermal
phonon bath and a radiation field. The interaction Hamiltonian can be written as $%
H_{I}=H_{p}+H_{r},$ with the phonon coupling
\begin{equation}
H_{p}=\sum_{m,n}^{N_{C}}q_{mn}^{p}a_{m}^{\dagger }a_{n},
\label{PhononBathHamiltonian-Eq}
\end{equation}%
where $q_{mn}^{p}$ is an operator acting in the bath Hilbert space.
The exciton\textit{-}phonon and exciton-thermal photon interaction
$H_{r}$ describes the coupling of the bath operators $q_{m}^{r}$ to
the transition
dipole moment of each chromophore, i.e.,%
\begin{equation}
H_{r}=\sum_{m}^{N_{C}}q_{m}^{r}\mathbf{(}a_{m}^{\dagger }+a_{m}).
\label{RelaxationHamiltonian-Eq}
\end{equation}%
The phonon terms $q_{mn}^{p}a_{m}^{\dagger }a_{n}$ induce relaxation
and dephasing without changing the number of excitations. In other
words, the state of the multichromophoric system remains in a fixed
excitation manifold under the evolution generated by $H_{p}$. The
$H_{r}$ Hamiltonian leads to transitions between exciton manifolds.
Generally, one can also consider diagonal static disorder in the
Hamiltonian as: $H_{d}=\sum_{m=1}^{N_{C}}\delta \epsilon
_{m}a_{m}^{\dagger }a_{m}.$ This disorder can be generated for
example by variation in the structure of the protein environment in
the time scales which are usually much slower than excitation
transfer time scale of $\sim 1$ ps \cite{Adolphs06}.

The dynamics of the system to second order in the system-bath
coupling, can be described by the Lindblad master equation in the
Born-Markov and secular approximations as \cite{BreuerBook}:

\begin{equation}
\frac{\partial \rho (t)}{\partial t}=-\frac{i}{\hbar
}[H_{C}+H_{LS},\rho (t)]+L_{p}(\rho (t))+L_{r}(\rho (t)).
\label{Master Eq}
\end{equation}%
where $H_{LS}$ are the Lamb shifts due to phonon and photon-bath
coupling.
The respective Lindblad superoperators $L_{p}$ and $L_{r}$ are given by\ ($%
k=p,r$):

\begin{eqnarray}
L_{k}(\rho ) &=&\sum_{\omega }\sum_{m,n}\gamma _{mn}^{k}(\omega
)[A_{m}^{k}(\omega )\rho A_{n}^{k\dagger }(\omega )
\label{Lindblad superoperator} \\
&&-\frac{1}{2}A_{m}^{k}(\omega )A_{n}^{k\dagger }(\omega )\rho -\frac{1}{2}%
\rho A_{m}^{k}(\omega )A_{n}^{k\dagger }(\omega )],  \nonumber
\end{eqnarray}%
For $H_{p}$ the corresponding Lindblad generators are
$A_{m}^{p}(\omega )=\sum_{\Omega -\Omega ^{\prime }=\omega
}c_{m}^{\ast }(M_{\Omega })c_{m}(M_{\Omega ^{\prime }})|M_{\Omega
}\rangle \langle M_{\Omega ^{\prime }}|$, where the summation is
over all transitions with frequency $\omega $ in the
single-excitation manifold and $|M_{\Omega }\rangle $ denotes the
exciton with frequency $\Omega $. The secular approximation is valid
when the relevant time scale of the intrinsic evolution of the
system, $\frac{1}{\left\vert \omega -\omega ^{\prime }\right\vert
}$, is much faster than the relaxation time scale. For example, an
energy difference of 200cm$^{-1}$ between excitons, e.g., in the FMO
complex, translates to a time scale of 0.16 ps which is much smaller
than the typical 1 ps time scale of energy relaxation in the
single-excitation manifold. The rates $\gamma _{mn}^{p}$ are given
by the Fourier transform of the bath correlation function as $\gamma
_{mn}^{p}(\omega )=\delta _{mn}\int dte^{i\omega t}\langle
q_{mm}^{p}(t)q_{mm}^{p}(0)\rangle ,$ where we assume that
off-diagonal fluctuations are small compared to diagonal
fluctuations \cite{Cho05}. The rate can be further simplified to the
site-independent expression $\gamma ^{p}(\omega )=2\pi \lbrack
J(\omega )(1+n(\omega ))+$ $J(-\omega )n(-\omega )]$ where $n(\omega
)=1/[\exp (\frac{\hbar \omega }{kT})-1]$ is the bosonic distribution
function at temperature $T$. Here, we assume an Ohmic spectral
density with\ $J(\omega )=0$ for $\omega <0$ and $J(\omega )=\frac{E_{R}}{%
\hbar }\frac{\omega }{\omega _{c}}\exp (-\frac{\omega }{\omega
_{c}})$ \ elsewhere, with cutoff $\omega _{c},$ and reorganization
energy $E_{R}=\hbar \int_{0}^{\infty }d\omega \frac{J(\omega
)}{\omega }$ \cite{Cho05}.

In the case of the Hamiltonian $H_{r}$ we obtain the Lindblad generators $%
A_{m}^{r}(\omega _{M})=c_{m}(M)|0\rangle \langle M|,$ where $\hbar
\omega _{M}=E_{Q_{\mathrm{Y}}}+\epsilon _{M}$ is the molecular
transition frequency, separated into $E_{Q_{\mathrm{Y}}}\sim
12500$\textrm{cm}$^{-1}$ for the $Q_{\mathrm{Y}}$ band of
bacteriochlorophyll $a$ and excitonic
energies $\epsilon _{M}$ of the order of 300cm$^{\text{-1}}$ \cite{Adolphs06}%
. The respective rate is again assumed to be diagonal, $\gamma
_{mn}^{r}(\omega )=\delta _{mn}$ $\gamma _{mm}^{r}(\omega ),$ and
site independent, $\gamma _{mm}^{r}(\omega )=\gamma ^{r}(\omega ).$ For Ohmic $%
\sim \omega $ and super-Ohmic spectral densities one
is able to approximate $\gamma ^{r}((E_{Q_{\mathrm{Y}}}+\epsilon
_{M})/\hbar )\approx \gamma ^{r}(E_{Q_{\mathrm{Y}}}/\hbar ).$ This
approximation yields a simplified Lindblad superoperator similar to
Eq.~(\ref{Lindblad superoperator}) without the summation over
frequencies and with the
generators $A_{m}^{r}=|0\rangle \langle m|.$ Finally, we choose\ the rates $%
\gamma ^{r}(\omega )$ such that it leads to a 1ns exciton life-time,
as experimentally measured for chromophoric complexes e.g., in Ref. \cite%
{Owens87}. The Lamb-shifts $H_{LS}=H_{LS}^{p}+H_{LS}^{r}$ are
explicitly given by $H_{LS}^{k}=\sum_{\omega ,n,m}S_{nm}^{k}(\omega
)A_{n}^{k\dagger }(\omega )A_{m}^{k}(\omega )$ (for $k=p,r$) where
$S_{nm}(\omega )$ is the imaginary part of the half-sided
Fourier-transform of the bath-correlator. The Lamb shift usually
contributes only marginally to the dynamics of the system, e.g. in
the FMO complex \cite{Adolphs06}.

The master equation Eq.~(\ref{Master Eq}) and its Lindblad
superoperators Eq.~(\ref{Lindblad superoperator}) contain a
significant degree of complexity, reflecting the non-trivial form of
the FMO complex and its diverse sources of environmental
interaction. To deal with this complexity one needs to separate the
contributions of the different physical processes to the dynamics.
To this end, in the next section, we explicitly construct a quantum
trajectory master equation to study the effects of free Hamiltonian,
damping and relaxation. Specifically, we demonstrate that the energy
transfer in multi-chromophoric systems of the chromophoric channel
can be considered as a generalized (directed) continuous-time
quantum walk in the single-excitation manifold interrupted by jumps
to the zero-excitation manifold.

\section{Quantum walk formalism for energy transfer}

\label{QW}

In general, the F\"{o}rster theory for energy transfer leads to a
classical
random walk description of the transport in photosynthetic units \cite%
{Sener02,Sener04}. The equation of motion for the classical
probabilities of an excitation being at site $a$, $P_{a},$ is given
by
\begin{equation}
\frac{\partial P_{a}(t)}{\partial
t}=\sum_{b=1}^{N_{C}}M_{ab}P_{b}(t). \label{Classical walk eq}
\end{equation}%
The $M_{ab}$ denotes the Markov transition matrix elements which
describes the classical F\"{o}rster rates between site $a$ and $b$.
In\ closely-packed chromophoric arrays, however, one has to consider
not only populations of states but also coherence between states;
consequently, the equation of
motion is given by a master equation for the density matrix such as Eq.~(\ref%
{Master Eq}). For the purposes of evaluating the dynamics of this
master equation, and for investigating the interplay between
coherence and decoherence in the resulting quantum walk, it is
convenient to re-express the master equation for a single excitation
in terms of a quantum trajectory picture of open quantum system
dynamics, as in \cite{Carmichael93,Castro07}:
\begin{eqnarray}
\frac{\partial \rho (t)}{\partial t} &=&-\frac{i}{\hbar
}[H_{eff},\rho
(t)]^{\star }  \label{quantumTrajectory-Eq} \\
&&+\sum_{m,m^{\prime },n,n^{\prime }}^{N_{C}}\Gamma ^{p}(m,m^{\prime
},n,n^{\prime })W_{m,m^{\prime }}\rho (t)W_{n,n^{\prime }}^{\dagger
}
\nonumber \\
&&+\sum_{m}^{N_{C}}\gamma _{m}^{r}R_{m}\rho (t)R_{m}^{\dagger },
\nonumber
\end{eqnarray}%
where the $W_{m,m^{\prime }}=a_{m}^{\dagger }a_{m^{\prime }}~$
generate jumps in the single-exciton manifold and the $R_{m}=a_{m}$
generate jumps between exciton manifolds. The jump rates are given
by$\ \Gamma ^{p}(m,m^{\prime },n,n^{\prime })=\sum_{l,\omega }\gamma
^{p}(\omega )\langle m|A_{l}^{p\dagger }(\omega )|m^{\prime }\rangle
\langle n|A_{l}^{p}(\omega )|n^{\prime }\rangle ,$\ and $\Theta
^{p}(m,n)=\sum_{l,\omega }\gamma ^{p}(\omega )\langle
m|A_{l}^{p\dagger }(\omega )A_{l}^{p}(\omega )|n\rangle $. We have
also defined $[,]^{\star }$
as a generalized commutation relation for any two operators $A$ and $B$ as $%
[A,B]^{\star }=AB-B^{\dagger }A^{\dagger }$. In order to describe
absorption of the exciton at the acceptor site and the damping due
to the phonon bath,
we introduced an effective anti-Hermitian Hamiltonian as: $%
H_{eff}=H_{C}+H_{LS}+H_{decoher}$, with
\begin{eqnarray}
H_{decoher} &=&-\frac{i}{2}\{\sum_{m,n=1}^{N_{C}}\Theta
^{p}(m,n)a_{m}^{\dagger }a_{n}  \label{DecoherHamiltonian-Eq} \\
&&+\sum_{m}\sum_{\omega }\gamma _{m}^{r}(\omega )a_{m}^{\dagger
}a_{m}+H_{C\rightarrow A}\}.  \nonumber
\end{eqnarray}

Now taking the trace of the overall master equation leads to the
probability density that no jump to the zero manifold occurs between
time $t$ and $t+dt$ as: $p_{no-jump}=-2i\mathrm{Tr}[H_{decoher}\rho
_{eff}(t)]dt+\sum_{m,m^{\prime },n,n^{\prime }}^{N_{C}}\Gamma
^{p}(m,m^{\prime },n,n^{\prime })\mathrm{Tr}[W_{m,m^{\prime }}\rho
_{eff}(t)W_{n,n^{\prime }}^{\dagger }]dt,$ and the probability
density of a
jump event becomes:\ $p_{jump}=\sum_{m}^{N_{C}}\gamma _{m}^{r}\mathrm{Tr}%
[R_{m}\rho _{eff}(t)R_{m}^{\dagger }]dt.$ In the case of a no-jump
trajectory one has the master equation $\frac{\partial \rho _{eff}(t)}{%
\partial t}=-\frac{i}{\hbar }[H_{eff},\rho _{eff}(t)]^{\star
}+\sum_{m,m^{\prime },n,n^{\prime }}^{N_{C}}\Gamma ^{p}(m,m^{\prime
},n,n^{\prime })W_{m,m^{\prime }}\rho _{eff}(t)W_{n,n^{\prime
}}^{\dagger },$ which can be considered as a directed quantum walk
on the one-exciton manifold described by the density operator $\rho
_{eff}$.\textbf{\ }

We note that the contribution due to the free Hamiltonian, i.e., the equation $%
\frac{\partial \rho (t)}{\partial t}=-\frac{i}{\hbar }[H_{C},\rho
(t)]$ represents a continuous-time unitary quantum walk in the
single-excitation manifold. The off-diagonal elements of the free
Hamiltonian terms represent quantum coherent hopping with amplitudes
$H_{C,mn}$ between the localized sites with energy $H_{C,mm}.$ The
analogy of continuous-time unitary quantum walks to classical random
walks was addressed within the context of quantum
algorithms in Ref. \cite{Farhi98} and further explored in \cite%
{Childs02,Kempe03}. The damping contribution to the dynamics within $-\frac{i%
}{\hbar }[H_{eff},\rho (t)]^{\star }$ leads to site-dependent
relaxation for both diagonal and off-diagonal elements of the
density operator and the terms $\sum_{m,m^{\prime },n,n^{\prime
}}^{N_{C}}\Gamma ^{p}(m,m^{\prime },n,n^{\prime })W_{m,m^{\prime
}}\rho (t)W_{n,n^{\prime }}^{\dagger }$ induce quantum jumps in the
single-excitation manifold.

The similarity between quantum walks and classical random walks, Eq.~(\ref%
{Classical walk eq}), can be readily emphasized in Liouville space.
The quantum trajectory picture is re-expressed as a matrix equation
by representing the density operator in a vectorial form. Thus,
quantum walks
in multi-chromophoric complexes are described by the equation%
\begin{equation}
\frac{\partial \vec{\rho}_{a}(t)}{\partial t}=\sum_{b}\mathcal{M}_{ab}\vec{%
\rho}_{b}(t),  \label{Quantum walk eq}
\end{equation}%
where $\vec{\rho}^{T}(t)=(%
\begin{array}{ccccc}
\rho _{11} & \rho _{12} & \dots & \rho _{N_{C}N_{C}-1} & \rho _{N_{C}N_{C}}%
\end{array}%
).$ This equation manifests a quantum walk in the
$N_{C}^{2}$-dimensional Liouville space, with the transition
(super-) matrix,
\begin{eqnarray}
\mathcal{M}_{ab} &=&-\frac{i}{\hbar }\{I\otimes
H_{eff}-H_{eff}^{\ast
}\otimes I)_{ab}  \label{TransitionSupermatrix} \\
&&+\sum_{m,m^{\prime },n,n^{\prime }}^{N_{C}}\Gamma ^{p}(m,m^{\prime
},n,n^{\prime })(W_{n,n^{\prime }}^{\ast }\otimes W_{m,m^{\prime
}})_{ab}. \nonumber
\end{eqnarray}%
The time variation of the vector $\vec{\rho}(t)$ in Liouville space
represents the dynamics of population elements $\rho _{mm}$ as well
as coherence elements $\rho _{mn},$ which are the signature of
quantum dynamics. The real part of $\mathcal{M}$ is responsible for
the directionality in the quantum walk which enhances the excitation
energy
transfer for the FMO complex as we will demonstrate later. This is due to the%
\textbf{\ }\emph{effective} interplay\textbf{\ }between the free
Hamiltonian and phonon-bath coupling that generates quantum jumps.
Next we study the properties of the quantum walk picture in terms of
the energy transfer efficiency as an universal measure.

\section{Energy transfer efficiency}
\label{ETE}

The energy transfer efficiency of the channel is defined as the
integrated probability of the excitation successfully leaving the
channel to the acceptor. This definition of the channel efficiency
has the advantage of being independent of the detailed dynamics
within the acceptor, charge separation, and the energy storage via a
chemical reaction. More formally we define:

\begin{equation}
\eta =\frac{1}{\hbar }\int_{0}^{\infty }\mathrm{Tr}(H_{C\rightarrow
A}\rho (t))dt.  \label{ETE eq}
\end{equation}%
\newline
Similar definitions,\ using \emph{integrated} success probabilities,
were
used in context of energy transfer from donor to acceptors \cite%
{Ritz01,Castro07}, and quantum random walks
\cite{Childs02,Blumen06}. Note that the efficiency in Eq.~(\ref{ETE
eq}) has the upper limit of $\infty ,$ yet in most practical cases
there is always a natural cutoff, $1/\gamma _{m}^{r},$ due to finite
excitation life-time, e.g., for chromophoric
complexes and GaAs quantum dots $1/\gamma _{m}^{r}\sim 1$ns \cite%
{Sener02,Nazir05}. Moreover, the relevant dynamical time scale for
the excitation to be transferred to the acceptor is about $1/\kappa
_{m}.$ A more quantitative measure for the transfer time through the
quantum channel
to the acceptor is given by,%
\begin{equation}
\tau =\frac{1}{\eta }\int_{0}^{\infty }t~\mathrm{Tr}(H_{C\rightarrow
A}\rho (t))dt.  \label{TransferTime eq}
\end{equation}

In this work we mainly focus on the energy transfer efficiency. In
order to study its optimality and robustness, we define a
decomposition $\Lambda _{k}$
of the superoperator by $\mathcal{M}=$ $\sum_{k}\Lambda _{k}.$The different $%
\Lambda _{k}$ can be chosen in any desired way in order to isolate
the effect of different parts of the master equation. For example,
one $\Lambda _{k}$ could represent the effect of the system
Hamiltonian on its own, while another could represent the effect of
the phonon bath. Each term $\Lambda _{k}$ is associated with a
scalar quantity $\lambda _{k}$, which represents the overall
strength of the of the $k$th term in the dynamics. For example, the
$\lambda _{k}$ associated with the Hamiltonian gives the overall
energy scale of the Hamiltonian, while the $\lambda _{k}$ associated
with the photon bath gives the strength of the coupling to that
bath.

It is natural to investigate the energy transfer efficiency in terms
of the susceptibilities $\frac{\partial \eta }{\partial \lambda
_{j}}$ and the
Hessian $\frac{\partial ^{2}\eta }{\partial \lambda _{j}\partial \lambda _{k}%
},$ using the scaling parameters $\Lambda _{k}\rightarrow $ $\lambda
_{k}\Lambda _{k},$ in the neighborhood of $\lambda _{k}\rightarrow $
$1.$ For any chromophoric complex the gradient of the efficiency
defined by the set of $\frac{\partial \eta }{\partial \lambda _{j}}$
can be used as a measure for local optimality of its performance
with respect to independent parameters $\lambda _{k}.$ Moreover,
such efficiency gradient can be utilized for engineering
photovoltaic materials with optimal and robust energy transfer in
their respective parameter space. One can verify that the
susceptibilities satisfy the identity $\sum_{k}\frac{\partial \eta }{%
\partial \lambda _{k}}=0,$ where the sum consists of a complete
decomposition of the superoperator. This identity is the consequence
of energy conservation for all times $t$: In the single-excitation
manifold, the energy that is transferred by the physical processes
$H_{C\rightarrow A}$ out of the chromophoric channel is absorbed by
the reaction center in a
given time interval. Additionally, the Hessian $\frac{\partial ^{2}\eta }{%
\partial \lambda _{j}\partial \lambda _{k}}$ is a measure for the
second-order robustness of natural or engineered chromophoric
systems. Note
that the robustness also satisfies a similar conservation property, i.e., $%
\sum_{j,k}\frac{\partial ^{2}\eta }{\partial \lambda _{j}\partial
\lambda_{k}}=0.$
%These conservation properties can be verified using the identity $%
%\eta =-\mathrm{Tr}(\mathcal{H}_{C\rightarrow A}\mathcal{M}^{-1}\vec{\rho}%
%(0)).$
Next, we study the quantum walk approach in the context the
ETE in the FMO complex.

\section{Fenna-Matthews-Olson complex}

\label{FMO}

\begin{figure*}[tbph]
%Use smaller figure for the arxiv
\includegraphics[scale=0.9]{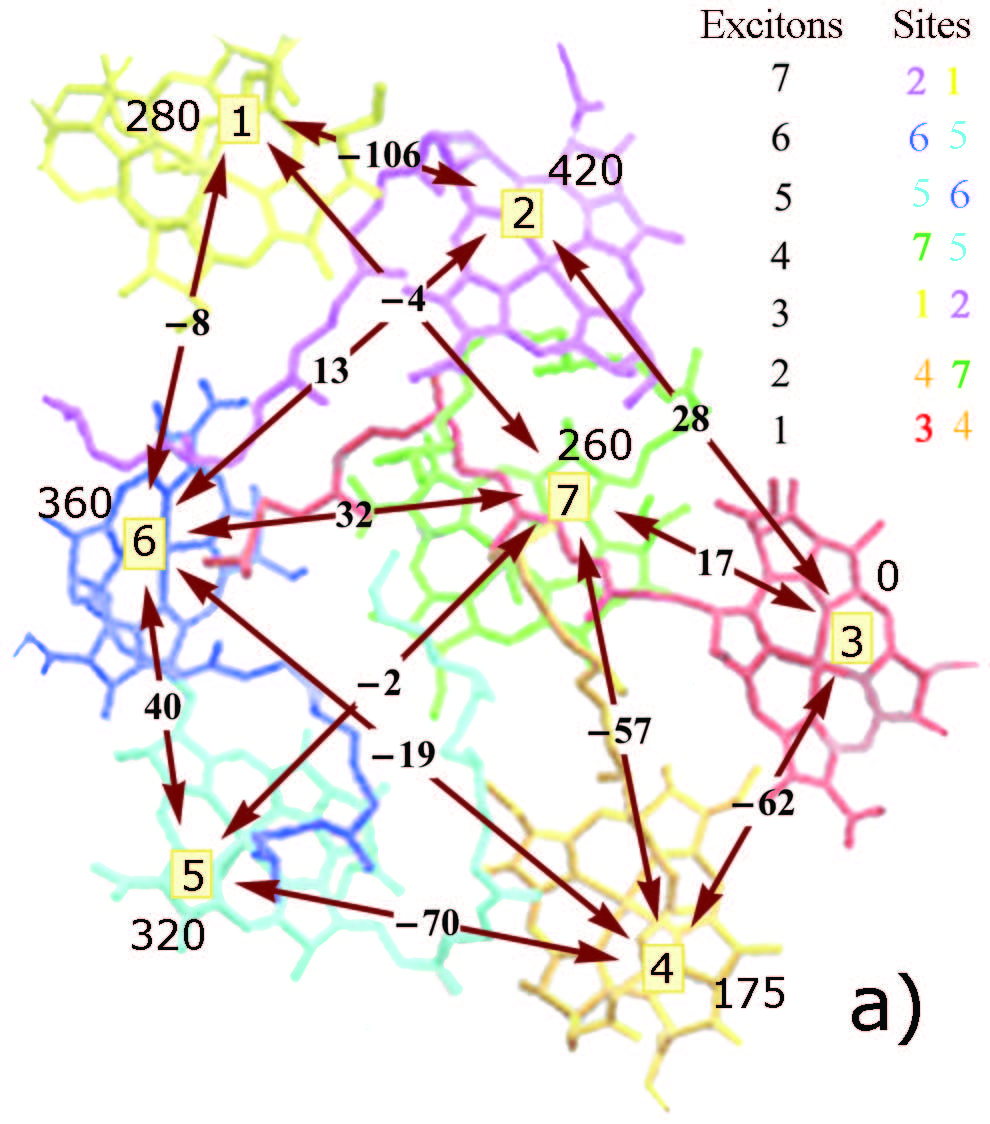}
\includegraphics[scale=0.9]{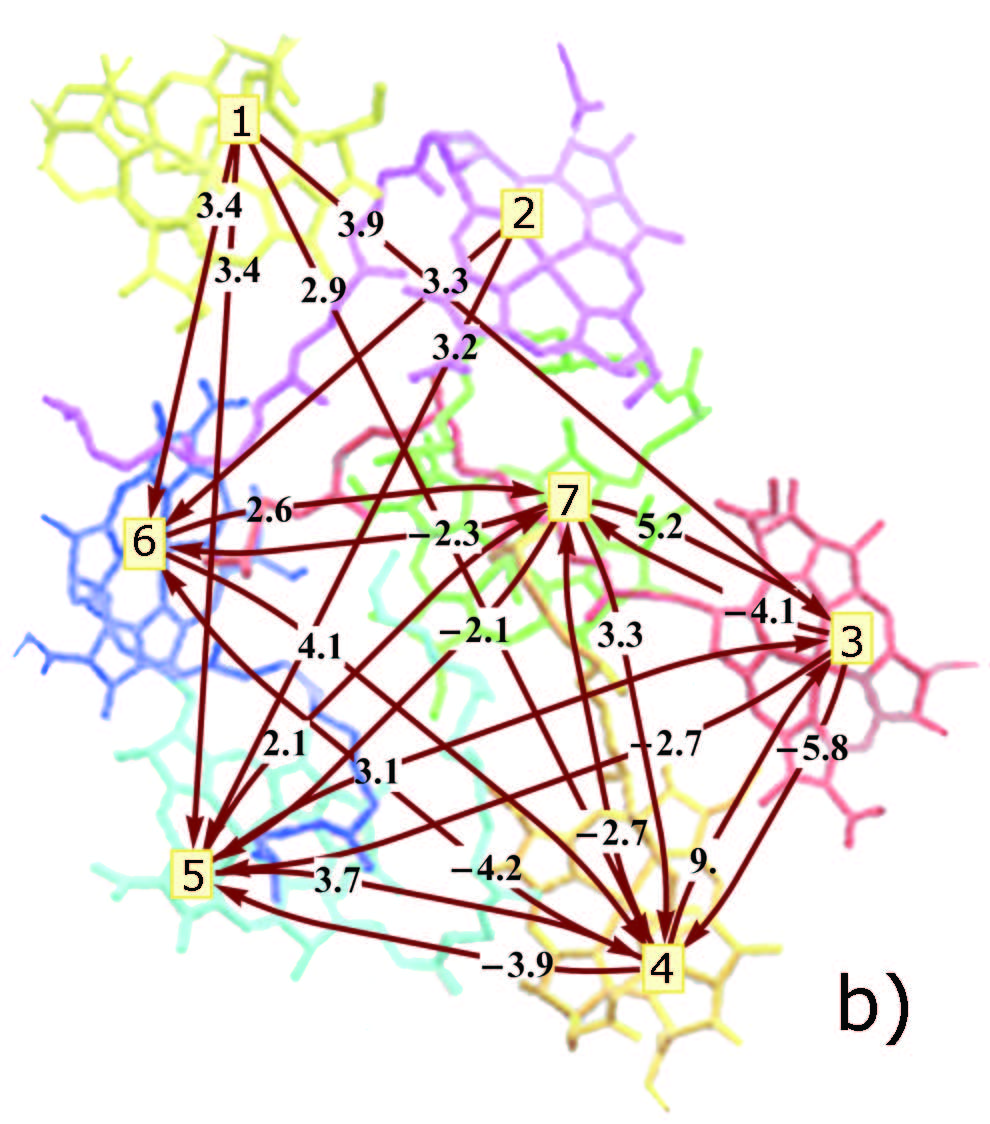}
\caption{The Fenna-Matthews-Olson protein: (a) The spatial structure
and energy levels of the complex, where the number at each site
represents the localized site energy and the arrows with numbers denote the
couplings among various bacteriochlorophylls. For clarity, some small couplings
are not shown. The inset depicts the participation of the seven chlorophylls
in the delocalized excitonic states \protect\cite{Cho05}.
(b) The susceptiblities of energy transfer efficiency with respect
to perturbations of inter-site jumps and corresponding damping,
rescaled by a factor of 10$^{4}$ and drawn with a cutoff of 2.0.
The initial state is taken to be a
mixture of populations at site 1 and 6. Standard
parameters are $E_{R}=35$cm$^{-1}$, $T=295$K, $\protect\kappa _{3}=1$ps$%
^{-1} $, and $\protect\gamma ^{r}=1$ns$^{-1}.$  Susceptibilites are
large when inter-chromophoric couplings are strong and site-energies
are similar. The sign of the susceptibility is an indication of the
directionality towards the target site 3. }
\label{fig3}
\end{figure*}

The structure of the Fenna-Matthews-Olson (FMO) complex of green
sulfur bacteria was revealed by x-ray crystallography \cite{Li97},
as the first pigment-protein complex structure to ever be determined
in this method, and since then it has been extensively studied
\cite{Mueh07}. The FMO complex consist of a trimer, formed by three
identical monomers, each constituting of seven bacteriochlorophyll
molecules (BChl a) supported by a rigid protein backbone. The FMO
complex essentially acts as a molecular wire, transferring
excitation energy from the chlorosomes, which are the main
light-harvesting antennae of green sulfur bacteria, to the
membrane-embedded type I reaction center. In the recent study of the
\textit{Chlorobium tepidum} FMO complex by Engel \textit{et al.},
\cite{Engel07}, using two-dimensional electronic spectroscopy,
direct evidence of long-lived coherence in the form of quantum
beatings was demonstrated at 77K. The presence of quantum coherence
prompted speculations about the presence of quantum computation in
FMO. Indeed, it was argued that FMO acts as a dedicated
computational device \cite{Engel07}, since excitons are able to
explore many states simultaneously and select the correct answer,
which here is the lowest energy excitonic state. This operation was
claimed to be \textit{analogous} to Grover's algorithm, which is
known to provide a quadratic speed-up over its classical
counterparts for searching elements of an unstructured database
\cite{Grover97}. Here, we argue that a purely unitary Grover-type
search algorithm cannot explain the efficiency of the exciton
transfer in FMO complex. However, we employ our directed quantum
walk approach to study the quantum effects in the dynamics of the
FMO complex.

We use the free Hamiltonian of the FMO complex as given in
Ref.~\cite{Cho05}. The site-energy differences and inter-site
couplings lead to exciton energies with separations of around
$100$cm$^{-1}$. The highest energy states are exciton 6 and 7 which
are mainly delocalized over sites 5/6 and 1/2, respectively. The
lowest exciton state 1 involves the site 3. For an overview of the
structure of the FMO complex see Fig.~\ref{fig3} (a). The initial
states for the simulation are taken to be sites 1 and/or 6 which are
close to the chlorosome antenna (donor) \cite{Adolphs06}. Transfer
of the excitation from the FMO channel to the acceptor occurs via
site 3 with the rate $\kappa _{3}$, which is a free parameter in our
simulations and, if not otherwise stated, taken to be $\kappa
_{3}=1$ps$^{-1}.$ The exciton lifetime is assumed to be $1/\gamma
^{r}=1$ns \cite{Owens87,Ritz01}. The bath spectrum is taken to be as
described above with the
reorganization energy $E_{R}=35$cm$^{-1}$ and cutoff $\omega _{c}=150$cm$%
^{-1},$ inferred from Fig.~2 of Ref.~\cite{Adolphs06}.

The purely unitary evolution generated by the seven-site Hamiltonian
is not performing a Grover-type search. In general, for a unitary
evolution to be qualified as a quantum search a certain set of
conditions have to realized: (i) $\left\langle\psi_{ES}\right\vert
\rho_{i}(t_{0})\left\vert \psi _{ES}\right\rangle =1-\alpha$, where
$\left\vert \psi_{ES}\right\rangle
=\frac{1}{\sqrt{N}}\sum_{m}\left\vert m \right\rangle$ is an equal
superposition of the basis states $\{\left\vert m \right\rangle
\}_{m=1}^{N}$ of the Hilbert space including the solution the
search, the target state $\left\vert m^{\ast} \right\rangle$.
$\rho_{i}(t_{0})=\left\vert \psi_{i}(t_{0})\right\rangle
\left\langle \psi_{i}(t_{0})\right\vert $ is the pure initial state
of the system at some arbitrary time $t_{0}$ and $\alpha\ll 1$. For
a standard Grover search algorithm $\alpha$ is equal to zero. This
condition assures that the initial input state of the search is
unbiased with respect to the target state. (ii) $\left\vert
\left\langle m^{\ast} \right\vert e^{-i(t_{f}-t_{o})H}\left\vert
\psi_{i}(t_{0})\right\rangle \right\vert ^{2}=1-\beta,$ where
$t_{f}$ is the smallest time in which the free evolution of the
initial state, $e^{-i(t_{f}-t_{o})H}\left\vert
\psi_{i}\right\rangle$, has significant overlap with the target
state, and $\beta\ll 1$.
(iii) $t_{f}=\gamma\sqrt{N},$ where $\gamma$ is either a constant or a $%
poly(\log N).$ In summary, a Hamiltonian generating the Grover
algorithm should map an equal coherent superposition of all possible
(database) states into a desired target state, within a time
polynomial in the size of database and with a probability of close
to one. For the FMO complex, we have investigated a variety of
different reasonable initial states (e.g., an equal superposition of
all BChl states, or a localized excitation on BChls 1 and 6) and
target states (e.g., the state localized in BChl 3). In neither case
did the coherent Hamiltonian dynamics result in a significant
overlap to the target state. We found that the overlap oscillates in
time but never exceeds the value of 0.4. We conclude that the
\emph{unitary} Grover search algorithm cannot explain the efficiency
of the exciton transfer in FMO complex. However, in principle,
certain non-unitary generalizations of quantum search algorithms
could still be developed to be relevant in this context; especially
for the case of non-unitary oracles interacting with a non-Markovian
and/or spatially correlated environment.

\begin{figure*}[tbph]
\includegraphics[scale=0.38]{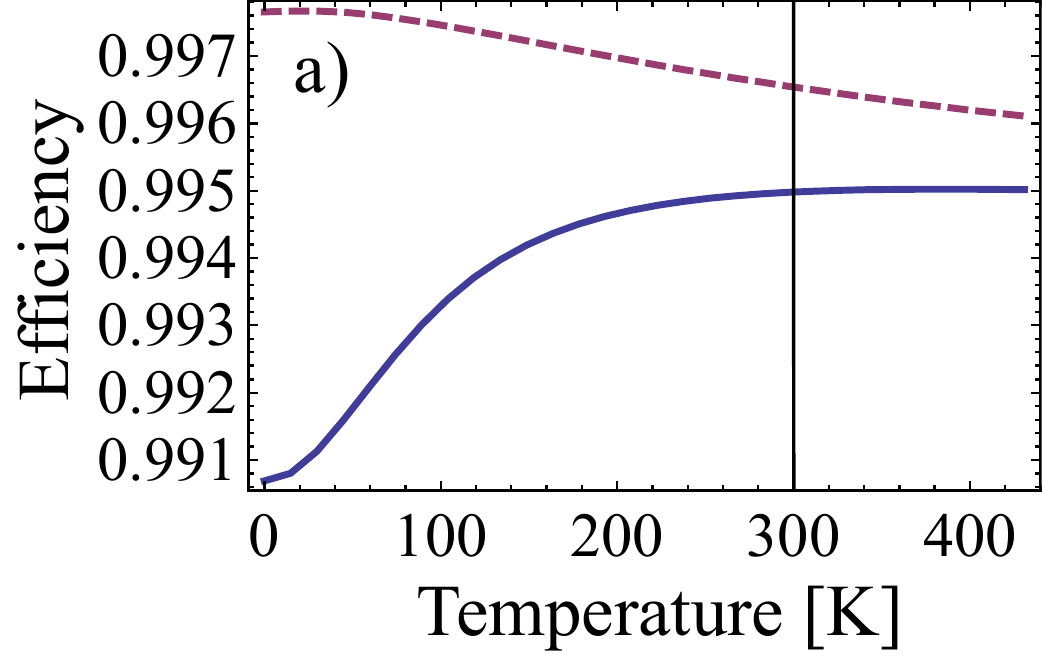}
\includegraphics[scale=0.38]{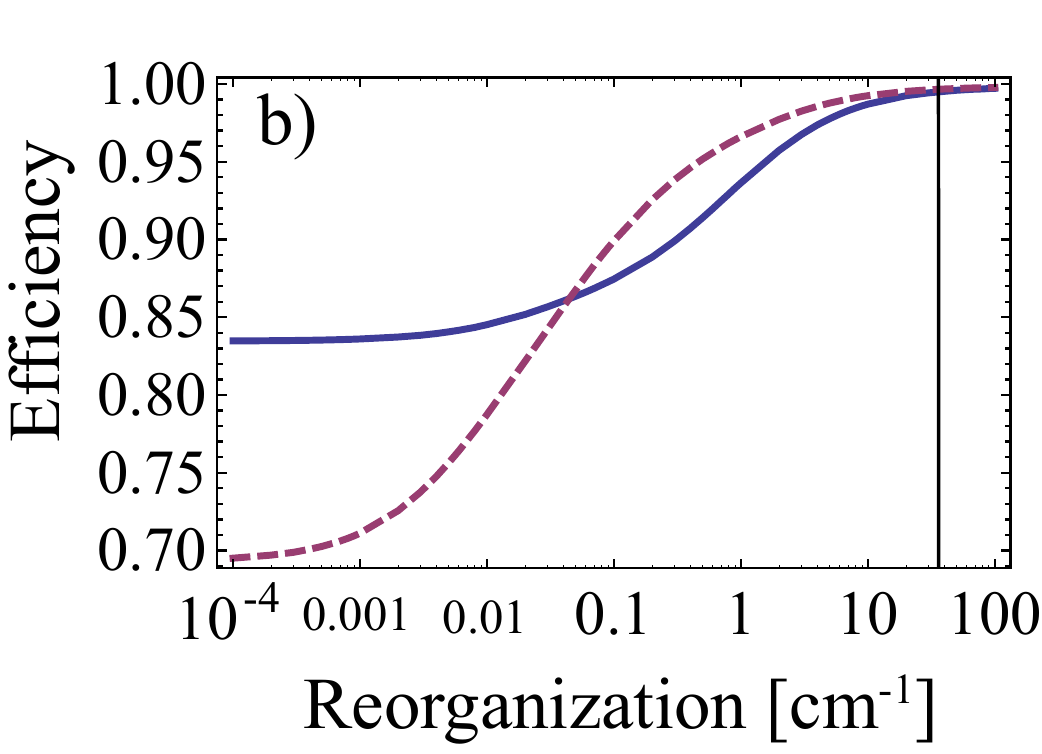}
\includegraphics[scale=0.38]{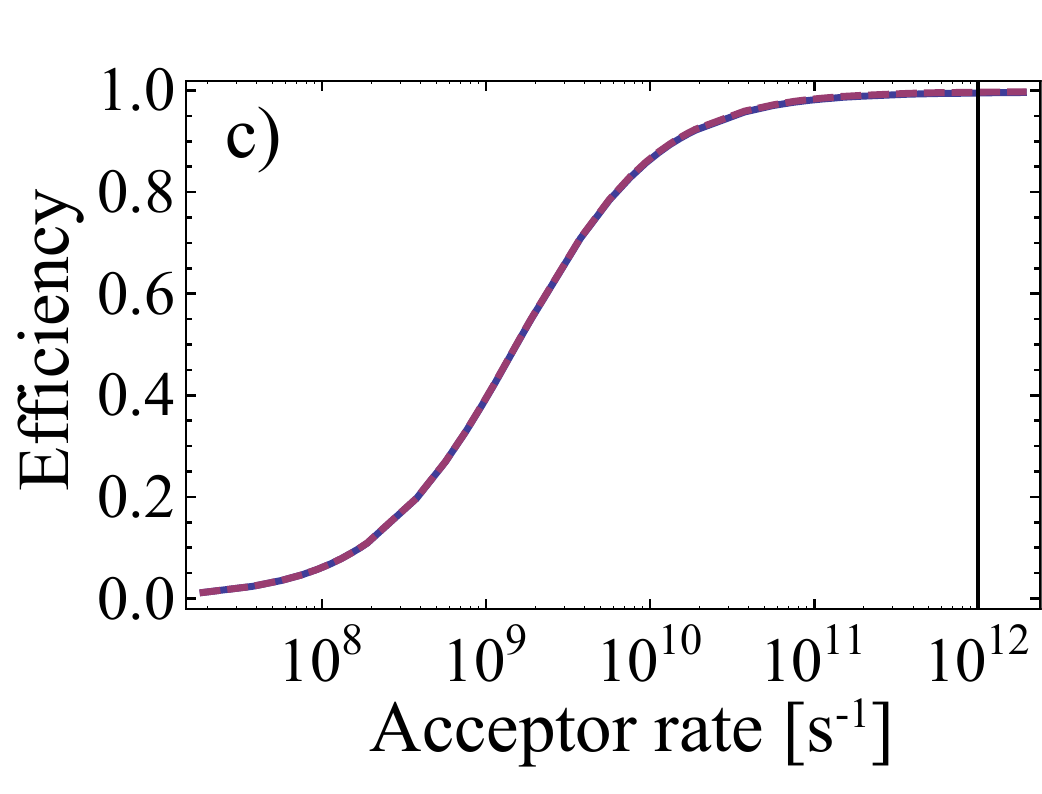}
\includegraphics[scale=0.38]{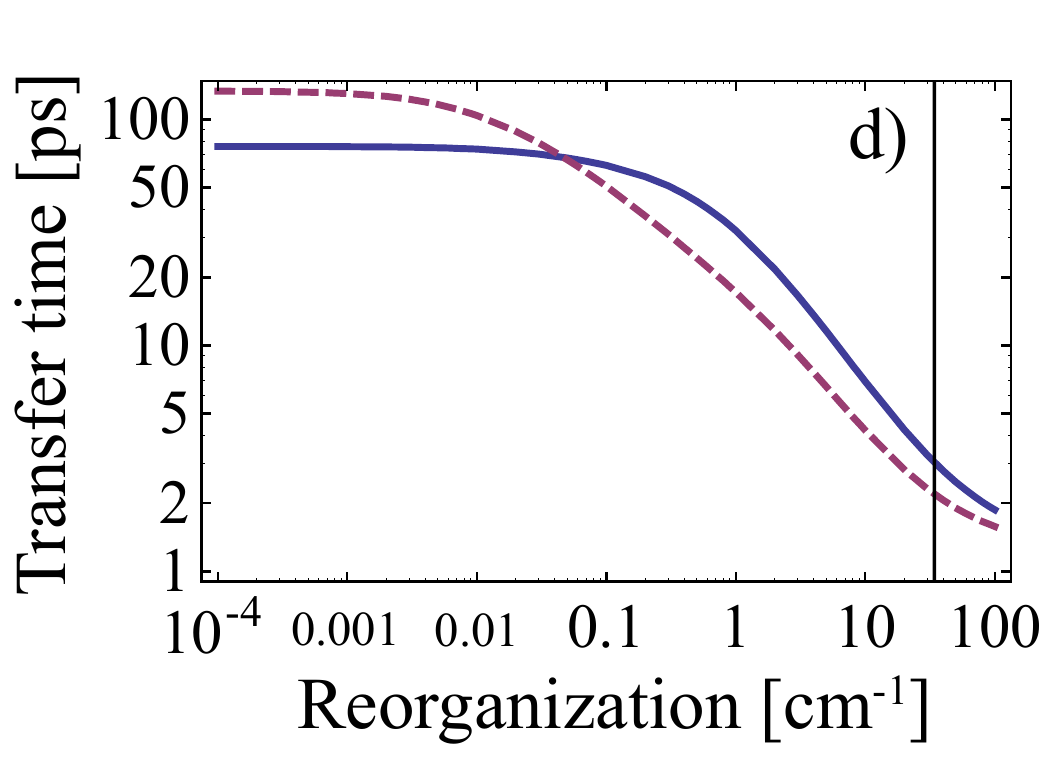}
\caption{ Energy transfer efficiency as a function of a)
temperature, b) reorganization energy (log-linear), and c) transfer
rate to the acceptor (log-linear). Transfer time as a function of
reorganization energy d). Blue lines show the efficiencies starting
from an initial state localized at site 1. Red lines how the
efficiencies starting from site 6. The default
parameters (shown as vertical lines) are taken to be $T=295$K, $\protect%
\kappa _{3}=1$ps$^{-1},$ $\protect\gamma ^{r}=1$ns$^{-1},$ and $E_{R}=35$cm$%
^{-1}$. A quantum walk with no environment-assisted jumps
corresponds to no reorganization energy in panel b). The energy
transfer efficiency in this case is 15-30\% less than for the
parameters obtained experimentally for FMO demonstrating the effect
of the environment-assisted quantum walk. } \label{fig1}
\end{figure*}

We investigate other non-trivial quantum dynamical effects in the
presence of phonon-bath fluctuations and exciton recombination and
trapping. The results are shown in Figs.~1, 2, and 3. In
Fig.~\ref{fig1} (a), we illustrate the functional dependence of the
ETE, Eq.~(\ref{ETE eq}), on temperature for two initial states
localized at site 1 or 6, respectively. The overall dependence is
less than 1\% for reasonable temperatures. This can be explained by
the relatively small size of the FMO and the approximately three orders of magnitude separation of lifetime ($%
1/\gamma ^{r}$) and acceptor transfer ($1/\kappa _{3}$) timescales.
In order to see this, note that at zero temperature there are only
quantum jumps originated from spontaneous emission of energy into
the phonon bath, leading to relaxation down the energy funnel. This
phenomenon in itself leads to a high efficiency of transport, due to
the presence of irreversible trapping on a time scale much faster
than the lifetime of the excitation. At higher temperatures quantum
jumps due to stimulated emission and absorption enter the dynamics.
Both processes have the same rates and a temperature dependence
which is determined by the bosonic distribution function
$n(\omega)$. In the FMO protein, stimulated emission of excitonic
energy helps the transport, since the target state has the lowest
energy. The effect of absorption is twofold: It facilitates the
overlap with the trapping site when there is an energy barrier in
the transport path. At the same time, absorption processes can lead
to transfer away from low-energy target sites (\textquotedblleft
detrapping\textquotedblright ). These effects explain the difference
in the temperature dependence for the two initial states localized
either at site 1 or at site 6. The ETE increases slightly with
temperature for the initial state being at site 1 (blue line in
Fig.~\ref{fig1} (a)). This site has a large overlap with exciton 3
and the spatial pathways to site 3 involve energetically higher
excitons 4,5,6, and 7, see Fig.~\ref{fig3} (a). Detrapping explains
the slight decrease of ETE as a function of temperature for initial
state 6 (red line in Fig.~\ref{fig1} (a)). This site has a high
overlap with exciton 5 and 6 and energetically funnels down to
exciton 1 (site 3). For larger photosynthetic complexes and/or in
the absence of low-energy trapping sites, the temperature dependence
of the efficiency is expected to be more significant. First,
temperature-independent spontaneous emission will play a less
prominent role. Second, the energy transfer time will become more
comparable to the exciton lifetime. This translates to a higher
temperature dependence for the overall efficiency since any
variation in the transport becomes more pronounced. mes.

The functional dependence of the ETE on the reorganization energy at
room temperature is shown in Fig.~\ref{fig1} (b). The reorganization
energy can be understood as a linear scaling of the phonon bath.
Thus, at $E_{\rm R}=0$ we have the (pure) quantum walk limit leading
to efficiencies of 70\% and 85\% respectively for both pathways. The
difference of 15\% between the two different initial states can be
explained by the higher localization of initial state 6 due larger
energy mismatch of site 6 and target site 3\textbf{. }An increase in
reorganization energy results in an efficiency increase up to about
99\%, demonstrating the effect of the environment-assisted quantum
walk.

As mentioned before the actual value of the irreversible transfer rate $%
\kappa _{3}$ from the FMO complex to the reaction center (acceptor) is not well known%
\textbf{,} since there exists insufficient chrystallographic data
for the combined FMO/RC structure \cite{Li97}. In Fig.~\ref{fig1}
(c), we explore the dependence of the ETE on this unknown parameter
at room temperature\textbf{. }The ETE increases monotonically from
zero to almost one within a range of five orders of magnitude. One
obtains the largest increase in efficiency when the acceptor
transfer rate is $\sim 1$ns$^{-1}$ thus surpasses the lifetime rate
$\gamma ^{r}.$ In the limit of large transfer rates to the acceptor,
$\kappa _{3}/\gamma ^{r}\gg 1,$ the lifetime of the excitation does
not significantly reduce the efficiency. On the other hand, in the
limit of small transfer rates to the acceptor, $\kappa _{3}/\gamma
^{r}\ll 1, $ most of the excitation dissipates into the environment
before being transferred to the acceptor.

Figure \ref{fig1} (d), shows the transfer time, Eq.~(\ref{TransferTime eq}%
), of the excitation initially at site 1 or 6 to the acceptor as a
function of reorganization energy. In the fully quantum limit it
takes the excitation more than 50ps to arrive at the acceptor. This
value dramatically improves for higher reorganization energies. At
the value of $E_{R}=35$cm$^{-1}$ one finds a transfer time of around
4ps$^{-1},$ which was reported based on different considerations in
\cite{Adolphs06}.

\begin{figure*}[tbph]
\includegraphics[scale=0.45]{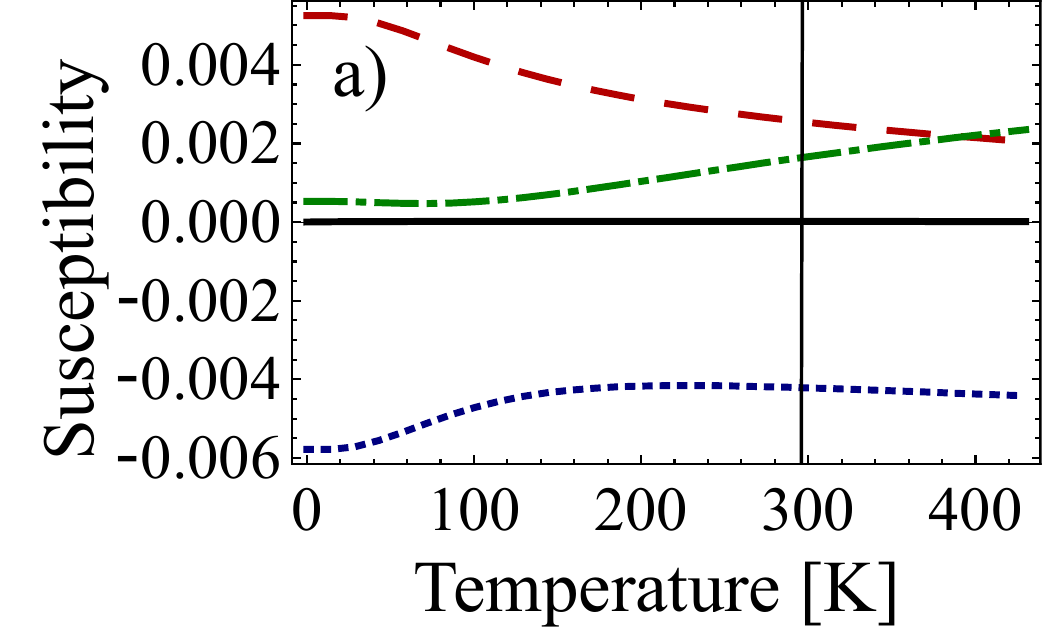}
\includegraphics[scale=0.45]{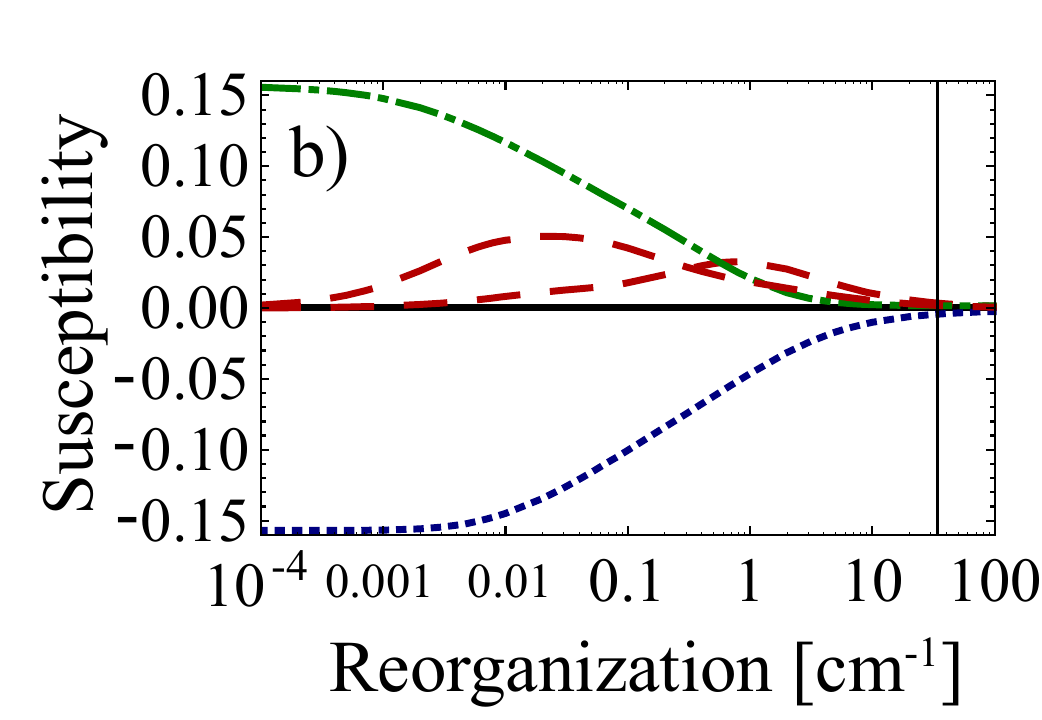}
\includegraphics[scale=0.45]{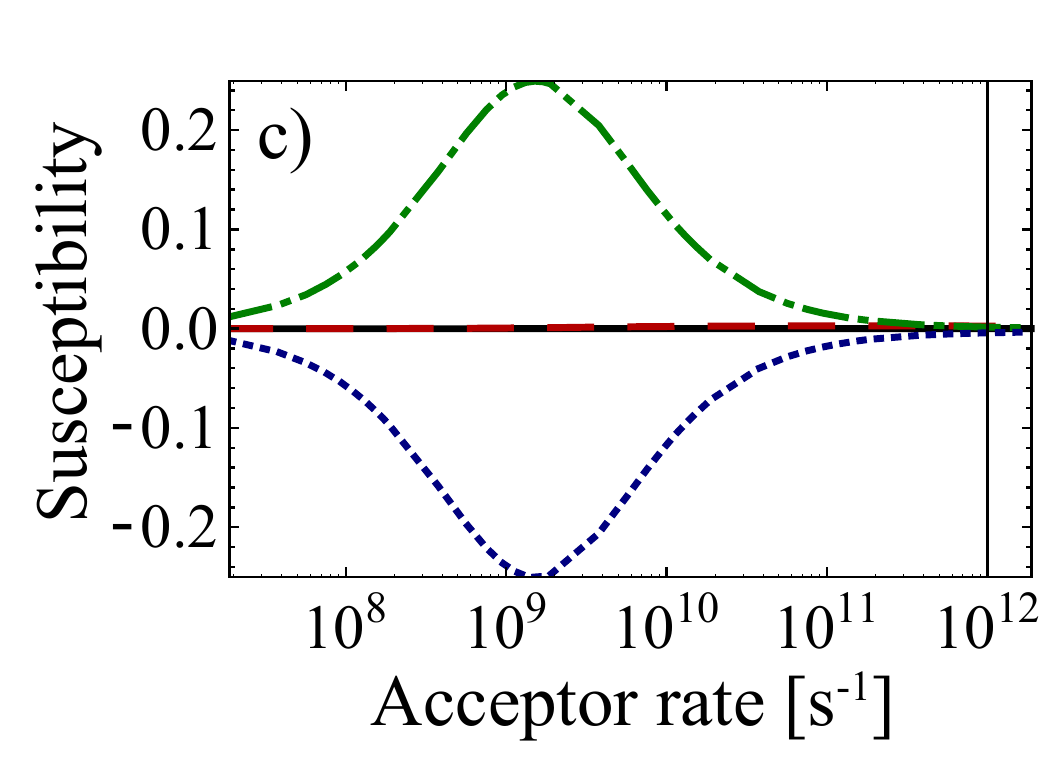}
\caption{Susceptibility versus a) temperature, b) reorganization
energy (log-linear), and c) transfer rate to the acceptor. The
processes considered are free Hamiltonian\ (black), the phonon bath\
coupling (red), transfer to the acceptor (green), and the loss of
the excitation (blue). The initial state is a mixture of 1 and 6,
except for the jump susceptibility in b) which is separated into 1
and 6 respectively. Standard parameters (shown as
vertical lines) are $E_{R}=35$cm$^{-1}$, $T=295$K, $\protect\kappa _{3}=1$ps$%
^{-1}$, and $\protect\gamma ^{r}=1$ns$^{-1}.$ For the standard
parameters, the FMO complex shows relatively small susceptibilities,
suggesting robustness with respect to external or evolutionary
perturbations.} \label{fig2}
\end{figure*}

Figure \ref{fig2} illustrates the susceptibilities $\frac{\partial \eta }{%
\partial \lambda _{j}}$ for the basic processes in the FMO dynamics
including the Hamiltonian, the phonon bath coupling, transfer to the
acceptor, and the loss of the excitation. In Fig.~\ref{fig2} (a),
the susceptibilities for those processes are shown as a function of
temperature. The susceptibility of the ETE to the phonon bath shows
several interesting features. The system is rather susceptible to
perturbations of the phonon bath coupling at zero temperature. In
this limit, the directionality of the quantum walk is maximized,
since, due to spontaneous emission, the excitation can only move
down in energy towards the target site 3 while energy absorption
from the phonon bath is suppressed. Increasing temperature leads to
increased stimulated emission \textit{and} absorption, thus the
system is less susceptible to perturbations of the phonon bath since
their effect on emission and absorption processes is the same.
Concomitantly, the efficiency becomes more susceptible to the
transfer process to the reaction center at higher temperature. This
is again readily explained by the increased phonon-bath absorption
rate: In the presence of detrapping processes it becomes more
important to capture the excitation immediately once it arrives at
site 3. We note that the efficiency is not susceptible to a small
variation of the free Hamiltonian due to inherent irreversibility of
the fully unitary quantum dynamics. The susceptibility with respect
to dephasing in the energy basis has also insignificant variation
for all the temperatures considered here due to the dominant role of
the quantum jumps to the overall ETE, and thus is not presented in
Fig.~\ref{fig2}.

The dependence of the susceptibilities of the ETE as a function of
the reorganization energy is shown in Fig.~\ref{fig2} (b). The
susceptibility of the ETE to the phonon bath peaks at around
1cm$^{-1}$ for the initial state 1 and around 0.01cm$^{-1}$ for the
initial state 6$.$ Note that, for the Ohmic spectral density
considered here, the reorganization energy scales linearly the
strength of the coupling to the phonon bath, therefore this curve is
the derivative of the ETE in Fig.~\ref{fig1} (b), where the steepest
ascent occurs around 1cm$^{-1}$ or 0.01cm$^{-1}$ respectively. The
susceptibility of the ETE on the transfer rate to the acceptor
becomes smaller for higher reorganization energy. Larger
reorganization energy leads to faster thermalization. In this regime
the system is more resilient to perturbations of the transfer
process to the acceptor and the dissipation to the environment.

The dependence of the susceptibility of the ETE on the acceptor
transfer rate is illustrated in Fig.~\ref{fig2} (c). We observe a
maximum in the susceptibility at around $1$ns$^{-1}$, which is the
regime where the time scales of transfer to the acceptor and the lifetime are comparable, $%
\kappa \sim \gamma ^{r}$, compare to Fig.~\ref{fig1} (c).

The susceptibilities of the energy transfer efficiency on inter-site
quantum jumps is shown in Fig.~\ref{fig3} (b). More formally, we
look at the effect of perturbations of site to site jump terms and the corresponding damping, $%
\Lambda _{nm}=-\Theta ^{p}(m,m)(I\otimes a_{m}^{\dagger
}a_{m}+a_{m}^{\dagger }a_{m}\otimes I)+\Gamma
^{p}(n,m,m,n)(W_{m,n}^{\ast
}\otimes W_{n,m}),$ in the quantum walk master Eq.~(\ref%
{TransitionSupermatrix}). We obtain a schematic picture of the most
susceptible pathways in the FMO complex. The susceptibilities are
correlated with the incoherent transport pathways. The
susceptibilities are large when relaxation or absorption of energy
due to the phonon bath is important for a particular transport
pathway. This is the case e.g.~when there is a relatively large
off-resonance between different sites and thus coherent coupling
does not lead to significant overlap between the sites, for example
for sites 3/7. On the other hand, when coherent coupling is similar
to the site-energy splitting, incoherent processes are less
significant, leading to small susceptibilities such as for sites 1/2
and 5/6. A positive (negative) susceptibility of a process
$\Lambda_{nm}$ indicates an increasing (decreasing) ETE. For the FMO
complex, jumps towards (away from) site 3 have positive (negative)
susceptibility, a signature of the directionality in the
irreversible dynamics. Thus, the quantum walk formalism together
with a rather straightforward measure of the energy transfer
susceptibility provides a valuable method to investigate spatial
exciton transfer pathways in multichromophoric arrays.

\section{Conclusions and future work}

\label{Conclusion}

We have developed a general theoretical framework within the
Lindblad formalism for studying the role of quantum effects in
energy transfer dynamics of arbitrary chromophoric arrays
interacting with a thermal bath from a microscopic Hamiltonian. We
have shown that a quantum walk approach, which has been widely used
in quantum information science, provides an appropriate mathematical
framework for studying energy transfer. We have generalized the
concept of continuous-time quantum walks to non-unitary and
temperature-dependent dynamics in Liouville space. This approach can
also be used to generally study decoherence effects in quantum walks
in arbitrary geometries. The energy transfer efficiency was used as
a universal measure to study the transfer properties of the
environment-assisted quantum walk. We We have applied our method to
explore the energy transfer efficiency and its susceptibilities for
the Fenna-Matthews-Olson protein complex as a function of \
temperature, reorganization energy, trapping rates, and quantum
jumps. In particular the energy transfer susceptibilities were
studied with respect to the free Hamiltonian, the phonon bath,
dephasing, trapping rate, and the exciton loss. This approach
provides valuable insight into the dynamical role of various
generators of the master equation and into spatial exciton transfer
pathways in multichromophoric arrays. We have shown that the overall
environment coupling strength leads to a substantial enhancement of
the ETE of about 25\% for the FMO complex. Thus, the overall energy
transfer efficiency of 99\% can be explained with the open nature of
the multichromophoric dynamics involving an effective interplay
between free Hamiltonian and fluctuations in the protein and
solvent.

We applied the general formalism presented in this work to reveal
the contributions of underlying physical mechanism to quantum
transport \cite{Rebentrost08-1}. We also quantified the phenomena of
environment-assisted quantum transport due to an effective interplay
of quantum dynamical coherence and a pure-dephasing noise model
\cite{Rebentrost08-2}. Using pure dephasing model, others observed
similar effects \cite{Plenio08-1,Strauch08}. Motivated by recent
observations of non-local effects in the structure and the dynamics
of the purple bacteria reaction center \cite{Lee07} and the FMO
protein \cite{Mueh07}, and also based on recent advances in the
understanding and control of non-Markovian open quantum systems
\cite{Rebentrost06,Mohseni08}, generalizations of our scheme to
include environments with temporal and spatial correlations are
currently underway. These results could lead to new ways for
engineering optimal state transfer in quantum spin networks
\cite{Christandl04,Plenio08-2} which interact with realistic
environments. Since we employ the Lindblad equation, our quantum
walk formalism can be implemented numerically using the Monte Carlo
wavefunction (MCWF)\ approach. The
MCWF method was originally developed for dissipative processes in quantum optics \cite{DalibardMolmer92}%
. The main advantage of MCWF\ is the fact that one only needs to
simulate the wave function rather than the density operator
\cite{Ohta06}. Our approach can potentially be used to enhance
energy transfer efficiency via engineering quantum interference
effects. For certain binary tree chromophoric arrays, e.g.,
dendrimers,\textbf{\ }quantum walks could lead to an exponential
speed-up over classical walks of excitations \cite{Childs02}. In
general, the combined biology and quantum information inspired
approach of this study could provide new insight for engineering
artificial photosystems, such as quantum dots and dendrimers
\cite{DendrimerReview} to achieve optimal energy transport by
exploiting their environmental effects.

%\begin{acknowledgments}
We would like to acknowledge useful discussions with J. Biamonte,
G.R. Fleming, I. Kassal, A. Najmaie, A.T. Rezakhani, and L. Vogt. We
thank the Faculty of Arts and Sciences of Harvard University, the
Army Research Office (project W911NF-07-1-0304) and Harvard's
Initiative for Quantum Science and Engineering for funding.
%\end{acknowledgments}

\end{document}